\documentclass[preprint2,times]{aastex6} 

\usepackage{graphicx}
\usepackage{times}
\usepackage{subfigure}
\usepackage{amsmath,amstext}
\usepackage{newtxmath} 
\usepackage{CJKutf8}
\usepackage[T1]{fontenc}

\newcommand{\kms}{\ensuremath{{\rm km~s}^{-1}}}

\newcommand{\Hoeq}{\ensuremath{H_0=70\,\kms \mbox{Mpc}^{-1}}}

\newcommand{\ha}{H$\alpha$}
\newcommand{\hb}{H$\beta$}
\newcommand{\hd}{H$\delta$}

\newcommand{\hdhost}{H$\delta_{\rm host}$}
\newcommand{\Msun}{M$_{\rm \odot}$}
\newcommand{\Mstar}{M$_\star$}
\newcommand{\fagn}{f$_{\rm AGN}$}

\slugcomment{Draft version Feb 28, 2018.}
\shorttitle{A Catalog of PSQ from SDSS DR7}
\shortauthors{Wei et al.}

\begin{document}
\title{A Catalog of Post-starburst Quasars from Sloan Digital Sky Survey Data Release 7}

\author{Peng Wei
\begin{CJK}{UTF8}{gbsn}
(魏鹏)
\end{CJK}
\altaffilmark{1,4}, 
Yang Gu
\begin{CJK}{UTF8}{gbsn}
(顾洋)
\end{CJK}
\altaffilmark{2},
Michael S.\ Brotherton\altaffilmark{3},
Yong Shi\altaffilmark{1,4} and 
Yanmei Chen\altaffilmark{1,4} }

\email{pwei.nju@gmail.com}

\altaffiltext{1}{School of Astronomy and Space Science, Nanjing University, Nanjing 210023, China}
\altaffiltext{2}{Nanjing Foreign Language School, Xianlin Campus, Nanjing 210023, China}
\altaffiltext{3}{Department of Physics and Astronomy, University of Wyoming, Laramie, WY 82071, USA}
\altaffiltext{4}{Key Laboratory of Modern Astronomy and Astrophysics (Nanjing University), Ministry of Education, Nanjing 210023, China}

\begin{abstract}

We present a catalog of nearby (z $\leq$ 0.5) quasars with significant
features of post-starburst stellar populations in their optical spectra, so-called
post-starburst quasars, or PSQs. After carefully decomposing spectra
from the Sloan Digital Sky Survey (SDSS) Data Release 7 (DR7) Quasar Catalog
into quasar and host-galaxy components, we derive a sample of 208 PSQs. Their host-galaxy components have
strong \hd\ absorption ($\rm EW \geq 6 \AA$) indicating a significant 
contribution of an intermediate-aged stellar population formed in a burst of
star formation within the past 1 Gyr, which makes them potentially useful for
studying the co-evolution of supermassive black holes and their host galaxies.

\end{abstract}


\section{Introduction}\label{sec:intro}

Supermassive black holes (SMBHs) exist at the centers of
essentially all massive galaxies \citep{Kormendy95}. 
Quasars, the most luminous Active Galactic Nuclei (AGNs), are powered by the accretion onto SMBHs.  
They are not only an important phase of SMBH growth, but also represent a key stage in the life cycle of massive galaxies (Heckman \& Best 2014).  
AGN feedback can terminate star formation in the host galaxy and mass accretion onto the SMBH \citep[][and references therein]{Fabian12}.

Theoretical studies have suggested two mechanisms responsible for triggering
starbursts and AGNs. First, in the early universe,
major-mergers ignite the the most luminous quasars and starbursts\citep[e.g.,][]{Bouwens09, Treister12}. Second, in recent epochs, 
the main fueling mechanisms may be entirely driven by secular processes
(e.g., Hopkins \& Hernquist 2009).

Recently, several studies have indicated that AGN and starburst activity may not be coeval. For example, \citet{Davies07} studied a sample of local AGN with IFU data, and found that strong AGNs are not present when the central stellar populations are of the order of a few 10 Myr old, implying a delay about 50­-100 Myr between the onset of central stellar mass growth and subsequent rapid black hole growth. Other recent studies  \citep[e.g., ][]{Schawinski09, Wild09, Wild10, Yesuf14} 
investigated AGN activity relative to the integrated stellar populations of local galaxies and suggested that AGNs become stronger after the stellar populations reach a few 100 Myr old.

Quasars with massive post­starburst hosts may also represent a vital intermediate phase before total star formation quenching, and may provide insight on the connection between AGNs and starbursts. Various observations lend support for the evolutionary scenario from ULIRGs to quasars to dead ellipticals \citep[e.g., ][]{Sanders88, Magnelli11, Murphy11}. A significant fraction of advanced­ merger (U)LIRGs (Guo et al. 2016 ) are obscured transiting post­-starburst galaxies \citep[see][]{Yesuf14},, which are the starting point of the fast evolutionary track.

The host galaxies of luminous AGNs are found to have intermediate­-age stellar populations \citep[e.g.][]{Canalizo13, Matsuoka15}, which may correspond to the stage after gas-rich mergers in the merger­ driven co-­evolutionary scenario. \citet{Zhang16} found that the specific star formation rates (sSFRs) decrease from ULIRGs to obscured 2MASS quasars \citep{Shi14} to unobscured PG quasars. On the other hand, infrared bright QSOs and narrow-­line Seyfert 1 (NLS1) galaxies are possibly in the early phase of the evolution from ULIRGs to dead ellipticals \citep{Hao05}.

Post­-starburst quasars (PSQs) simultaneously show the spectral signatures of quasars and post­-starburst stellar populations \citep{Brotherton99}. The observed spectra of
PSQs show the power-­law continuum and the broad emission lines, as well as strong Balmer absorption,characteristic of intermediate age stellar populations 
\citep{Brotherton99, Brotherton02}.

A small sample of luminous PSQs, spectroscopically selected from the SDSS at z $\sim$ 0.3 , was investigated in detailed using HST/ACS F606W imaging (Cales et al. 2011 ), high S/N Keck and KPNO optical spectroscopy \citep[][hereafter C13]{Cales13} and \textit{Spitzer} mid-infrared spectroscopy \citep{Wei13}.. The selection criteria of this PSQs sample used the Balmer absorption lines and the Balmer break of the quasar spectra in order to identify the post­-starburst signatures.

These aforementioned studies found that PSQs have a heterogeneous host population with different stellar properties (starburst mass and age) and AGN properties (SMBH mass and Eddington fractions). They concluded that the PSQs with early­-type hosts likely evolved via major mergers, while those with spiral hosts were triggered by secular processes.
In this work, we compile a large sample of PSQs based on careful AGN­-host decomposition. Large samples of post­-starburst quasars, covering large areas, have not yet been identified before . In the rest of this section we briefly review works relevant to post-­starbursts quasars.

\citet{Kauffmann03c}  found that a significant fraction of high-­luminosity type 2 AGN have experienced a starburst in the recent past. Likewise, \citet{Goto06} identified 840 type 2 AGNs with post­-starburst hosts from the SDSS, and found that the fraction of post-­starburst type 2 AGNs is at least 4.2\% of all galaxies in their sample.

Similarly, \citet{Matsuoka15} identified 191 type 1 quasars at z $< $ 1 using high Signal­-to­-noise ratio (SNR) optical spectra from the 7 deg$^2$ field of the SDSS Reverberation Mapping (RM) project. They derived host stellar properties such as age and mass, and found that half of the quasars in this sample show post-­starburst signatures.

\citet{Melnick15} compiled a sample of 72 nearby PSQs from the SDSS DR7Q catalog. Unlike \citet{Matsuoka15}, who used a spectral decomposition and measured EW(H$\delta$), \citet{Melnick15} directly measured the EW(H$\delta$) from the observed quasar spectra without estimating the dilution of this absorption line by the corresponding quasar emission line. In addition, their lenient (EW(\hd) $> 3 \AA$) cut cannot reject large number of quasars which are hosted by normal star-­forming galaxies.

This work is different from preceding works for the following reasons: 1) Our work focuses on the post­-starburst quasars, which are brighter than $M_{i} < -22$ mag. In contrast, several large sample studies only focused on post­ starburst hosts of the low­-luminosity type 2 AGNs \citep[cf.][]{Kauffmann03c, Goto06, Yesuf14}. 2) We use the spectral decomposition analysis of our spectra to account for the dilution of post­-starburst features by the AGN component. Several previous investigations did not include such analysis \citep[cf.][]{Cales11, Cales13, Melnick15}. 3) We applied a stricter criterion of (EW(\hd) $> 6 \AA$)  to select the pure post­-starburst hosts with larger post­-starburst stellar populations fractions and effectively reject the normal star-­forming hosts \citep[cf.][]{Goto06, Melnick15}. 4) The parent sample of this work has larger areal coverage than that of SDSSRM quasar sample \citep[cf.][]{Matsuoka15}. Our spectral decomposition method is similar to that of \citet{Matsuoka15}.

Our sample of quasars and the selection criteria for PSQs are described in
\autoref{sec:sample}. Details of the spectral fitting method are explained in
\autoref{sec:fitting}. We test our method in \autoref{sec:test}. 
Results and
discussion are given in \autoref{sec:results} \& \ref{sec:discuss}. At last,
we summarize our results in \autoref{sec:summary}. We will discuss a more
detailed study of the properties of PSQs in a subsequent paper. We adopt
cosmological parameters \Hoeq, $\Omega_{m}=0.3$, and $ \Omega_{\Lambda}=0.7$
throughout this paper.

\section{Sample Selection and Fitting Method}\label{sec:sample_and_fitting}

For our sample of PSQs, we begin with the SDSS DR7 Quasar Catalog
\citep[DR7Q,][]{Schneider10}, which consists of the 105,783 SDSS quasars with
$M_{i} < -22$ mag and at least one broad optical emission line with $\rm FWHM
\geq 1000$ \kms. The absolute magnitude limit in DR7Q catalog was
calculated  by correcting the BEST i-band PSF magnitude measurement for Galactic
extinction  (using the maps of Schlegel et al. 1998) and assuming that the
quasar spectral  energy distribution in the ultraviolet-optical can be
represented by a  power-law with $\alpha = -0.5$ \citep[see details
in][]{Schneider10}. 

Our PSQ sample inherits this absolute magnitude cut from the parent DR7Q sample. This absolute magnitude cut is close to the classical quasar
threshold $M_B = -23mag$ (Schmidt \& Green 1983), which corresponds to  $M_i
\sim -23 mag$ \citep{Matsuoka15}. However, both AGN and host can  contribute to the
i-band PSF luminosity. For an observed PSQ spectrum,  the fraction of light
contributed by the host galaxy must be large enough to show their spectral
signatures. Due to this compound nature of  PSQs, this absolute magnitude cut  ($M_{i} < -22 mag$) means our sample is not complete down to a specific AGN luminosity or a
stellar population  luminosity.

\begin{figure*}[!htb]
	\centering 
	\subfigure[]{ 
		\includegraphics[width=0.45\textwidth]{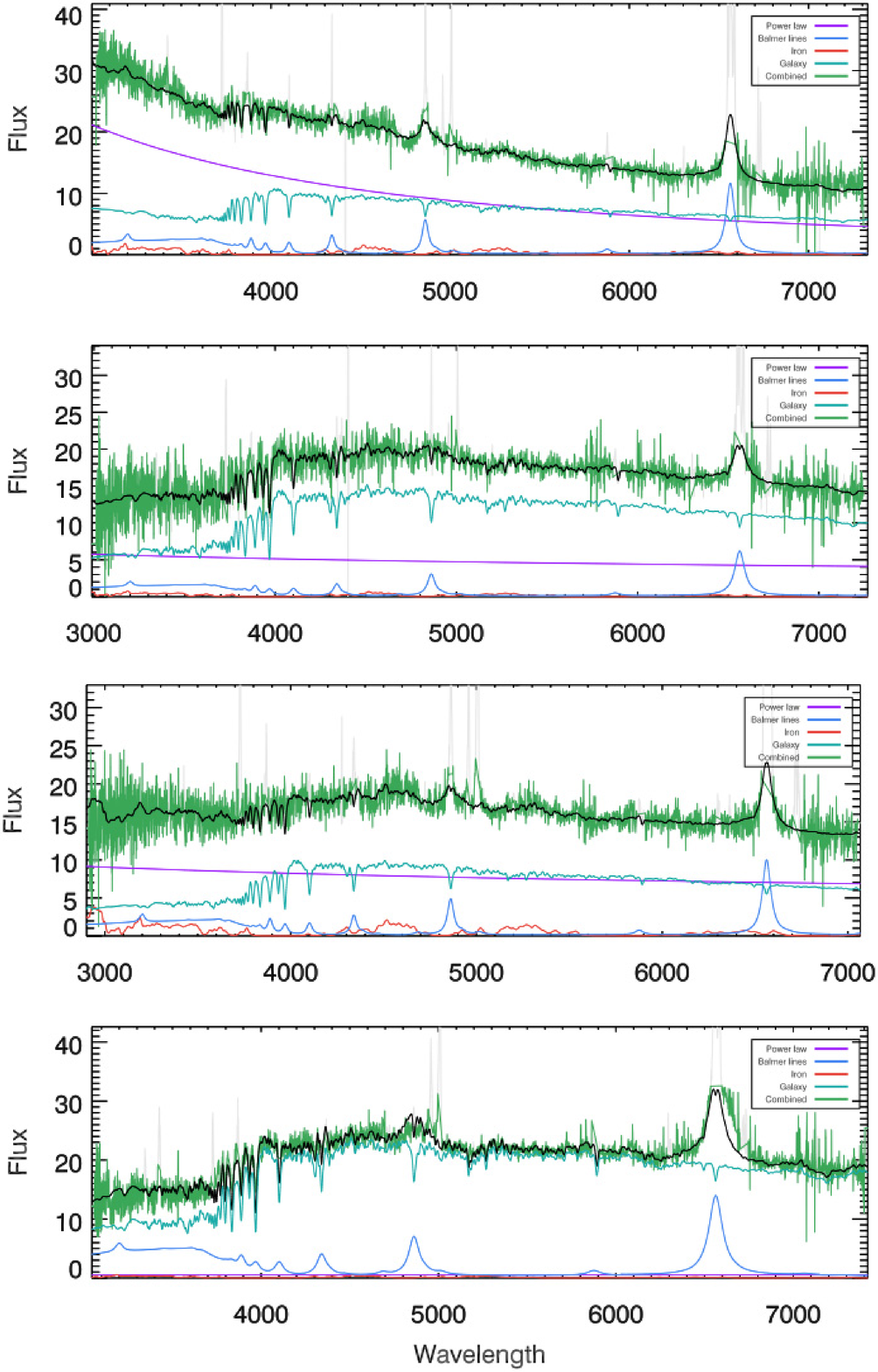}}
	\subfigure[]{ 
		\includegraphics[width=0.445\textwidth]{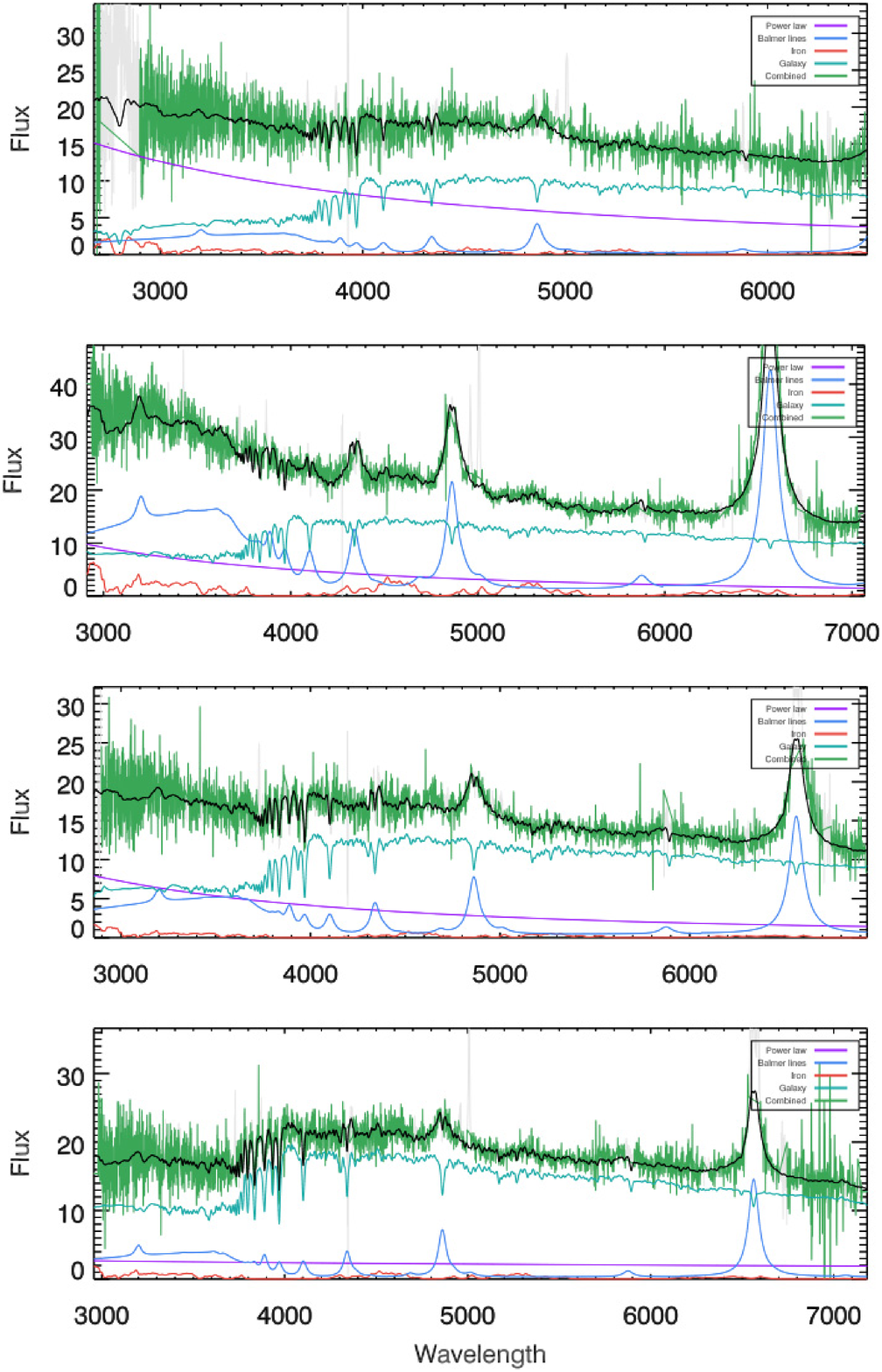}}

	\caption{\label{fig:fitting} 
	Examples of the spectral decomposition. The black
	line indicates the observed spectra, while the dark green line shows the
	total model spectrum. The other lines represent the individual model
	components, namely, the quasar power-law continuum (purple), the Fe II
	emission (red), the gas emission from the BLR (blue), and the stellar
	emission (light green).}
\end{figure*}

\subsection{Sample selection}\label{sec:sample}

Strong Balmer absorption arises in galaxies that experienced a recent burst of star
formation triggered $0.1 - 1$ Gyr previously with subsequent rapidly quenching.
After a major starburst epoch has ceased, the aging population of the
starburst will produce a characteristic post-starburst spectral signature with
strong Balmer lines in absorption. The EW of the \hd\ 
absorption line, one of the star-formation 
history indicators, allows us to constrain mean stellar ages of
galaxies and the fractional stellar mass formed in bursts over the past few
100 Myr.

However, unlike the selection of post-starburst galaxies or post-starburst type 2 
AGNs, a selection criterion based on the Balmer absorption line of observed quasar 
spectral depends both on the age of the stellar populations as well as the amount 
of quasar dilution. Hence, for the most reliable measurements of the host 
components (\hdhost), it is important to apply spectral decomposition methods. 

The initial selection criteria for the spectra for which we apply spectral decomposition are:

\begin{itemize}
	\item $S/N_{4150 \sim 4250\AA}$ $\geq$ 10
	\item z $<$ 0.5
	\item EW(\hd) $\geq -2 \AA$
\end{itemize}

$S/N_{4150 \sim
4250 \AA}$ is the continuum signal-to-noise ratio between the rest-wavelengths
of 4150 and 4250 \AA. This S/N criterion requires that spectra have enough S/N
to decompose robustly, and approximately corresponds to $m_{i} \leq 18.6mag$.

We limit the redshift range to z $<$ 0.5 and reject higher redshift quasars
because they have luminous AGN contributions and
weak stellar features, making them hard to decompose. That is
also the reason why we did not use a later SDSS data release, which mainly
focus on high-luminosity, high-redshift quasars. For example, a 
z = 0.5 quasar with $m_{i}  =18.6mag$ (i.e. $S/N \sim 10$) has an absolute 
magnitude  $M_{i}  = -23.5mag$. If the host is a massive galaxy with $M_{i}  
= -22mag$, there is only 25\% light contributed by the host which is 
only 2.5 times the noise. 

The EW(\hd) used for this initial step is the direct measurement of the total observed quasar spectrum. EW(\hd) $\geq -2\AA$  
ensures that the broad Balmer emission line from the quasar 
BLR is not too strong to hide the Balmer absorption line from the host. 

There are 1682 quasar spectra that satisfy the above criteria in the SDSS 
DR7Q catalog. We performed fitting and decomposition (see next section) of the 
spectra of those quasars. We exclude 438 objects with the fraction of light 
contributed by AGN ($f_{AGN}$) larger than 75\%, 
because the light contributed by the host may be comparable to the noise, which 
makes the spectral fitting results unreliable. After a careful check by 
eye, 80\% (987/1244) spectra have successful spectral fitting and decomposition.

For classification as a post-starburst quasar, one additional criterion from the 
decomposition result below has to be fulfilled: EW(\hdhost) $\rm \geq 6 \AA$.
According to the tests in appendix A, EW(\hdhost) is the most robust
parameter of the host component that we can recover with our spectral decomposition. It
is pointed out that, although our decomposition provides quantitative
measurements of mean stellar age, in the rest of this paper we use
 EW(\hdhost) as the primary diagnostic parameter. 

\begin{figure}[!htb]
	\center{\includegraphics[width=0.45\textwidth]{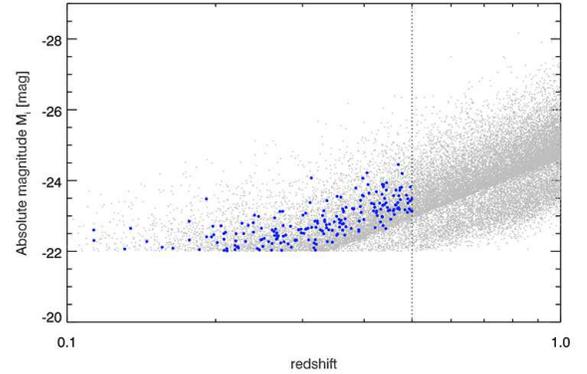}}
	\vspace*{2mm}
	\caption{\label{fig:mi_z} 
	Redshifts and i-band absolute magnitudes of the DR7Q quasars (small dots) and 208 post-starburst quasars (large dots). The dotted line is our redshift cut (z $<$ 0.5) in this study.}
\end{figure}

We found that about 21\% of the objects (208
out of 987) have rest-frame equivalent withs of \hdhost\
greater than 6\AA\ in absorption, which is the final PSQs sample as listed in \autoref{tab:table1}.
\autoref{fig:mi_z} displays the distribution of their redshifts 
and absolute magnitudes in SDSS i band ($M_i$). 
Our PSQ fraction is significantly
higher than the 4.2\% post-starburst AGNs studied by
\citet{Goto06} in a volume-limited sample of SDSS galaxies, where they derive the
fraction of all AGN with strong \hd\ in absorption, while our sample is
restricted to objects classified as quasars (\fagn $\leq$ 75\%). In our sample, 
25 PSQs meet the criteria of \citet{Melnick15}, \textbf{and 6 PSQs are in the 38 PSQs sample of \citet{Cales13}}.

\citet{Kauffmann03c} find that 95\% of galaxies with EW(\hd) $\rm > 6\AA$
 have experienced a burst with a mass fraction greater than 5\%  during the last
2 Gyr. If using a more relaxed \hd\ 
absorption criterion (i.e. EW(\hd) $ > 4$ \AA), it is difficult to distinguish
post-starbursts from normal star-forming galaxies with constant SFR.
Therefore, the selection post-starburst galaxies also uses a limit on nebular
emission to restrict the amount of on-going star formation. 
However, \citet{Falkenberg09} found this approach to
be too narrow to cover the full range in post-starburst populations. 
Recent investigations \citep[e.g.,][]{Wild09, Yesuf14} have identified the 
precursors of post-starburst galaxies have shown that AGNs 
are more common in these objects but that there is a significant time lag 
between the starburst and the AGN phase. In fact,
while requiring nebular lines to be weak only identifies objects in the quenched post-starburst \citep[QPSB]{Yesuf14} phase, but overlooks the quenching post-starburst galaxies in
transit between the starburst stage and the fully QPSB stage
Furthermore,
limits on nebular emissions also exclude any post-starburst galaxies hosting
AGNs since AGNs can power nebular emission lines. Given these issues,
the non-negligible number of AGNs found in post-starburst
 galaxies should cause us to reconsider the
relative importance of the presence of an AGN in post-starbursts. We should
not be too surprised that only 1\% of all galaxies are quenched post-starbursts 
galaxies \citep{Wong12}, because of the
underestimation of the transiting post-starbursts with ongoing star formation
or AGN activity .

We therefore did not place limits on [OII]$\lambda3727$ or \ha\ emission. 
The only constraint we used is that
the equivalent width of \hd\ in absorption is EW(\hdhost) $\geq$ 6 \AA. 
It ensures that high A-star fraction in
the stellar population dominating the luminosity. This is similar to
\cite{Yesuf14} who invented plausible criteria for identifying
transiting post-starbursts, as well as \cite{Bergvall16}.
There may be some more contemporary starburst in our PSQs without limits on the
[OII] or \ha\ lines. From the figure 6 of \citet{Kauffmann03a}, we can see there 
still exist starburst components at EW(\hd) $> 6 \AA$. \textbf{In future work}, we will 
use [OII]/[OIII] to gauge the ionization level, and use mid-infrared data and 
the SED of PSQ to estimate the SFRs of PSQs, and to determine the fraction of 
starbursts in our sample.

\begin{deluxetable*}{cccccccccccc}
\tabletypesize{\scriptsize}
\tablecaption{The properties of Post-starburst Quasars. \label{tab:table1}}
\tablewidth{0pt}
\tablehead{
\colhead{SDSS DR7 NAME} &
\colhead{RA (2000)} &
\colhead{DEC (2000)} &
\colhead{$z$} &
\colhead{$i$} &
\colhead{$M_i$} &
\colhead{\fagn} &
\colhead{\hd} &
\colhead{\hdhost} \\
\colhead{} &
\colhead{(deg)} &
\colhead{(deg)} &
\colhead{} &
\colhead{(mag)} &
\colhead{(mag)} &
\colhead{} &
\colhead{(\AA)} &
\colhead{(\AA)} \\
\colhead{(1)} &
\colhead{(2)} &
\colhead{(3)} &
\colhead{(4)} &
\colhead{(5)} &
\colhead{(6)} &
\colhead{(7)} &
\colhead{(8)} &
\colhead{(9)}  
}
\startdata
000104.92+160101.2& 0.27&	16.02&	0.45&	18.97&	-23.36&	0.60&	0.71&	6.21 \\
002815.78+004247.5& 7.07&	0.71&	0.31&	18.27&	-23.30&	0.39&	5.93&	9.45 \\
002959.03+150817.2& 7.50&	15.14&	0.21&	18.04&	-22.40&	0.57&	0.00&	7.52 \\
021652.47-002335.3& 34.22&	-0.39&	0.30&	18.29&	-23.31&	0.53&	-1.24&	7.66 \\
031715.10-073822.3& 49.31&	-7.64&	0.27&	18.32&	-22.93&	0.51&	0.74&	8.32 \\
032628.53-002741.4& 51.62&	-0.46&	0.45&	18.75&	-23.72&	0.65&	0.90&	8.50 \\
072837.76+443646.0& 112.16&	44.61&	0.28&	18.35&	-23.07&	0.27&	2.66&	6.92 \\
074310.24+461014.5& 115.79&	46.17&	0.27&	17.99&	-23.32&	0.21&	2.65&	6.38 \\
074621.06+335040.7& 116.59&	33.84&	0.28&	18.08&	-23.26&	0.23&	3.54&	6.66 \\
074844.87+441422.0& 117.19&	44.24&	0.43&	18.72&	-23.51&	0.41&	4.40&	9.17 \\
...
\enddata
\tablecomments{
This table is available in its entirety in a machine-readable form in the
online journal. A portion is shown here for guidance regarding its form and
content.}
\end{deluxetable*}

\subsection{Spectral-fitting method}\label{sec:fitting}

In order to disentangle the AGN contribution from that of the host galaxy, we
performed a spectral fitting and decomposition. We
utilized the IDL MPFIT to model the host stellar populations and AGN
contributions to the DR7 SDSS spectra. The $\chi^{2}$ minimization
technique of MPFIT simultaneously models multiple components:


\begin{multline}\label{eq1}
	S(\lambda)= aA(\lambda,\sigma_{Fe})+bB(\lambda,\sigma_{B})\\
	+cC(\lambda,\alpha_{\lambda})+\sum_{i=1}^6 d_{i}~SSP_{i}(\lambda,\sigma_{*})
\end{multline}

$S(\lambda)$ is the observed spectrum. $A(\lambda,\sigma_{Fe})$ denotes the
UV and optical iron emission line blends, which are modeled using the UV and
optical Fe II templates derived from I Zw 1 \citep[]{Vestergaard01,
Boroson92}. We convolved a Gaussian of width $\sigma_{Fe}$ with the Fe II
templates to simulate different velocity widths. $B(\lambda,\sigma_{B})$
represents the templates of Balmer emission lines and Balmer continuum
broadened by convolving with a Lorentzian of width $\sigma_{B}$ with a specific
Balmer decrement. The AGN power-low continuum is assumed to be
$C(\lambda)=\lambda^{\alpha_{\lambda}}$. The slope $\alpha_{\lambda}$ is
assumed to extend from $-$2.5 to 0, which approximately reproduces the observed range
\citep{Shen11}. Finally, $\sum_{i=1}^6 d_{i}~SSP_{i}(\lambda,\sigma_{*})$
represents the starlight component modeled by the 6 Simple Stellar Populations
templates with the solar metallicity and different age (30 Myr, 100 Myr, 300
Myr, 1 Gyr, 3 Gyr and 10 Gyr), which had been built up from the spectral
template library of \citep[hereafter BC03]{BC03}. These were broadened by
convolving the spectra with a Gaussian of width $\sigma_{*}$ to match the
stellar velocity dispersion of the host galaxy.

The fitting was performed by minimizing $\chi^{2}$ with $\sigma_{Fe}$,
$\sigma_{B}$, $-\alpha_{\lambda}$, $\sigma_{*}$, $a$, $b$, $c$, and $d_{i}$
being non-negative free parameters.  We give plots of the run corresponding
to the best fit of first eight objects as examples in \autoref{fig:fitting}.

We note that we did not employ younger SSP templates (e.g., $<$ 30 Myr) 
because they may be degenerate with 
the quasar power-law continuum. We emphasize that we only use our fitting to approximately reconstruct the host spectra. This fitting does not do detailed modeling of the stellar populations of the hosts. A more sophisticated model may use more SSP templates of varying ages and metallicities than we did. 


The narrow emission lines from the interstellar medium (ISM) and the AGN
narrow-line region (NLR) are masked out during the fitting. Our default models do
not incorporate the effects of dust extinction. The amount of intrinsic
extinction observed in unobscured quasars is usually small \citep{Richards03,
Hopkins04, Salvato09, Matute12} and its effect is limited. For half of our 
sample, the rest-frame spectral range doesn't cover blueward of the Mg II 
line, because of the redshift distribution of our sample. Hence we did not fit the 
Mg II line. 
 
\section{Results}\label{sec:results}

\textbf{We constructed a catalog of 208 PSQs. Most of them are identified as such for the 
first time here.} We used spectral decomposition of AGN and stellar population 
of the host galaxies. We defined post-starbursts as host galaxies with EW(\hdhost) 
$\rm \geq 6\AA$. Since we are interested in PSQs as a class as well as their
typical characteristics, we derived composite optical spectra and composite SEDs.

\begin{figure*}[!htb]
	\center{\includegraphics[width=0.8\textwidth]{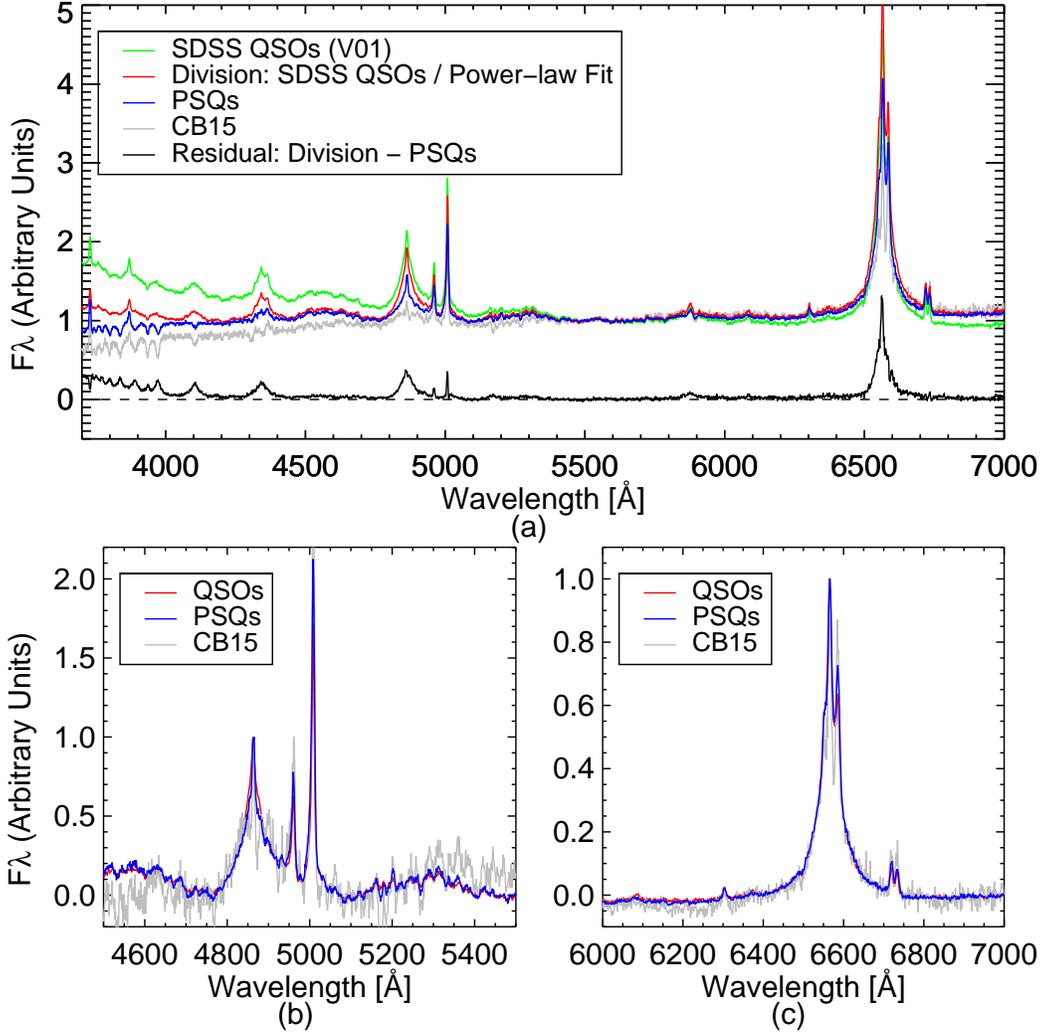}}
	\caption{\label{fig:compsp} 
	(a) Composite optical spectra of PSQs in our sample, that of \citet{Cales15} 
	and that of SDSS QSOs \citep{Vanden.Berk01}. They are normalized at 5500 \AA. 
	The red line indicates the SDSS QSOs spectra which was removed the power law 
	continuous, while the green line shows the total model spectrum. (b) The detailed 
	view of \hb\ region. The the red line shows the composite spectrum of PSQs (blue) 
	and the blue one shows SDSS QSOs (red). They are 5100\AA\ continuum subtracted
 and are normalized at the peak flux of \hb. (c) The detailed view of \ha\ 
	region, and the colors are assigned similarly to (b). The 6800\AA\ continuum 
	flux is subtracted and the spectrum are normalized at the peak flux of \ha.}
\end{figure*}

\begin{figure}[!htb]
	\center{\includegraphics[width=0.5\textwidth]{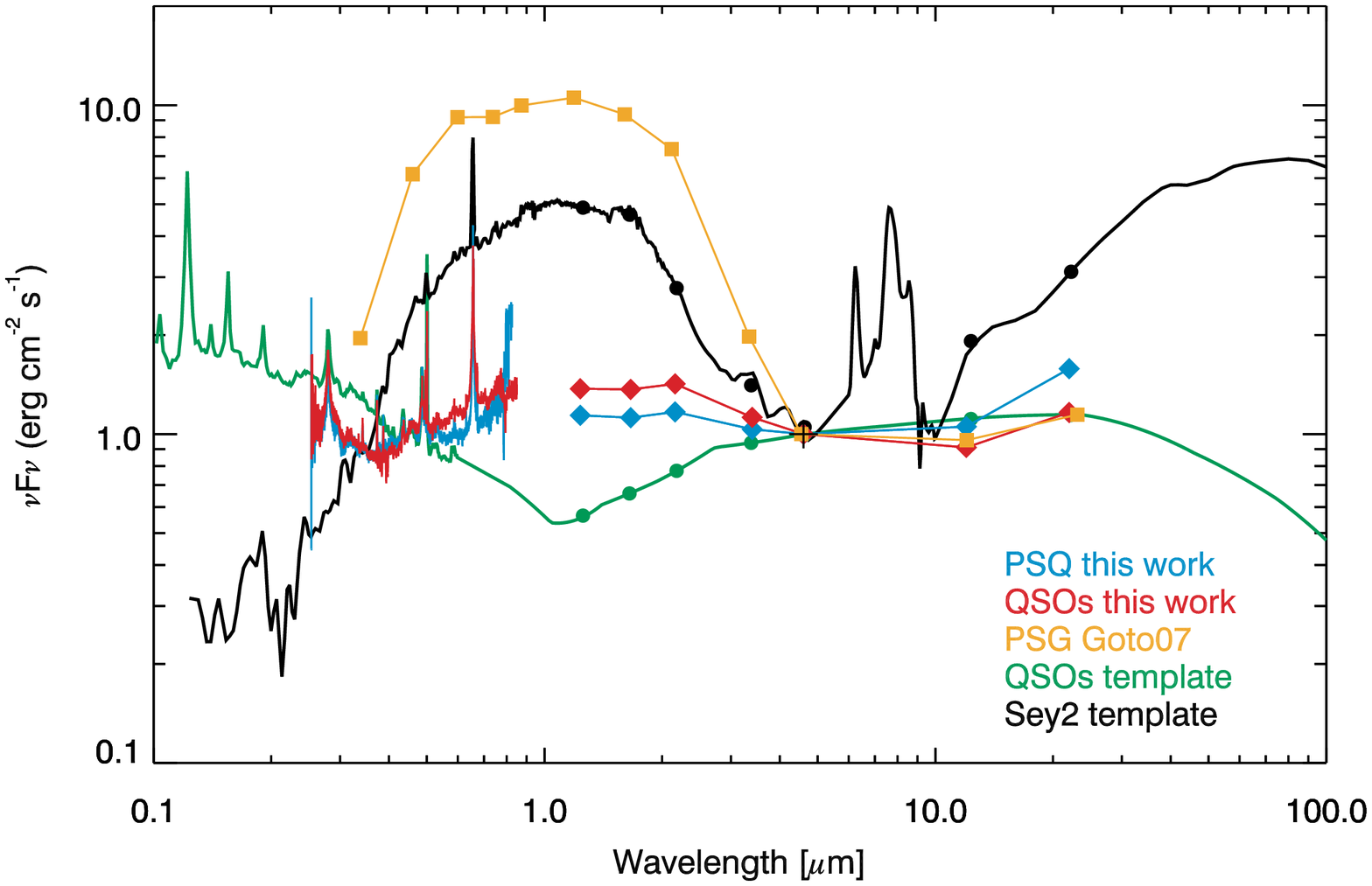}}
	\caption{\label{fig:compsed} 
	The composite SEDs in $\nu F_{\nu}$ vs $\lambda$ of our full quasar sample
	and PSQ subsample, compared with the SED template of Seyferts 2,
	QSOs (SWIRE Template Library; Polletta07), and that of post-starburst galaxies
	 \citep{Goto07}, normalized at the \textit{WISE} 4.6\micron\
	W2 band.}
\end{figure}

\subsection{Composite spectra of PSQs}\label{sec:compsp}

We computed the composite optical spectra of the total sample of 208 PSQs by 
 rebinning the spectra to the rest frame using accurate redshifts, normalizing the 
 spectra at rest-frame 5500\AA, and applying a median stacking. A previous study 
 \citep[][hereafter CB15]{Cales15} also created composite spectra of PSQs, but the 
 AGN luminosity and post-starburst criteria are different between our sample and 
 theirs. In \autoref{fig:compsp}, we compare our composite spectrum of PSQs with 
 that of SDSS QSOs \citep[][hereafter V01]{Vanden.Berk01} and that of CB15.

Our PSQs spectrum has a relatively flat continuum compared to V01. In order 
to better compare our spectrum to the V01 spectrum, we normalized the V01 spectrum by dividing out its power-law 
continuum. Our spectrum is similar to the normalized V01 and CB15 spectra at the red end. At 
the blue end, we can see the obvious the Balmer high-order absorption lines in our PSQ 
spectrum similar to the spectrum of CB15, which shows the post-starburst characteristics, 
unlike V01 which only has Ca K absorption line. 
We can also 
see the Fe emission features in our spectrum. The \ha\ and \hb\ profiles are very similar 
to those of V01, which indicates that they have similar widths and asymmetries of broad 
Balmer emission, similar ratio of narrow Balmer line to broad Balmer line, and the ratio 
of SII to \ha.

\subsection{Composite SED of PSQs}\label{sec:sed}

We also derived the composite SED of the PSQ sample, where the
optical photometric data are from SDSS DR7, the near-ultraviolet
(\textit{NUV}) magnitudes are from the \textit{GALEX} satellite
\citep{Martin05}, near-infrared magnitudes are from 2MASS, and the 
mid-infrared magnitudes are from \textit{WISE}.

We created the composite SED of the PSQs by normalizing their spectra at rest-frame
4.6\micron\ and applying a median combine at rest-frame wavelengths.
We repeated the procedure for the quasar sample. In \autoref{fig:compsed}, we
show composite SED of PSQs and the SED templates of Seyferts 2, QSOs (SWIRE
Template Library; \cite{Polletta07}), and post-starburst galaxies
\citep{Goto07}.

It is clear that the PSQs lie between Seyfert 2 and QSO SEDs in the mid-infrared, 
especially between 3.4\micron\ and 12\micron\ where they are closer to
QSOs. The bumps at optical and near-infrared wavelengths are due to a
significant stellar fraction  in our PSQs. The flat mid-infrared
SED of post-starburst galaxies is very similar to our quasar and QSO
templates, which can be fitted with a hot dust component plus an AGN model
\citep{Sajina12}. The 22\micron\ excess of PSQs might correspond to the dust
heated by young stars created during ongoing star formation.

\section{Discussion}\label{sec:discuss}

\subsection{Modeling of composite SEDs}\label{sec:sed_modeling}

We performed the detailed modeling of the composite SEDs in \autoref{fig:compsed} with 
an AGN SED fitting code to determine the age of stellar component and the properties of 
the star-forming. The code is AGNfitter \citep{Calistro16}, an
open-source code allows the user to disentangle components responsible for their 
emission robustly through a fully Bayesian Markov Chain Monte Carlo algorithm to 
fit the SEDs of AGNs. A large library of theoretical, empirical, and semi-empirical 
models are used to characterize both the AGN and its host radiation simultaneously. 
Four physical radiation components constitute the model: stellar populations, cold dust 
in star-forming regions, an emitting accretion disk and a torus of hot dust surrounding 
the AGN. AGNfitter allows the user to derive the integrated luminosities, dust 
attenuation parameters, stellar masses, and SFRs, by calculating the posterior 
distributions of numerous parameters governing AGNs' physics through a fully Bayesian 
treatment of errors and parameter degeneracies. We use the mean value of the probability 
density function (PDF) as the final value of each parameter (see \autoref{tab:table2}), 
while the 16th and 84th percentiles of the distribution give the associated uncertainty.

\begin{deluxetable*}{cccccccc}
\tabletypesize{\scriptsize}
\tablecaption{Comparisons of fundamental physical properties \label{tab:table2}}
\tablewidth{0pt}
\tablehead{
\colhead{Type} &
\colhead{PDF percentiles} &
\colhead{$\tau$} &
\colhead{Age} &
\colhead{\Mstar} &
\colhead{$\rm L_{IR(8-1000)}$} &
\colhead{$\rm SFR_{IR}$} \\
\colhead{} &
\colhead{} &
\colhead{(Gyr)} &
\colhead{(log yr)} &
\colhead{(log \Msun)} &
\colhead{($erg\ s^{-1}$)} &
\colhead{(\Msun $\ yr^{-1}$)} \\
\colhead{(1)} &
\colhead{(2)} &
\colhead{(3)} &
\colhead{(4)} &
\colhead{(5)} &
\colhead{(6)} &  
\colhead{(7)}  
}
\startdata
			& 16\%&	1.83 &	7.96 &	10.32 &	40.74 &	0.00 & \\
PSQs	& 50\%&	5.69 &	8.74 &	10.58 &	44.19 &	6.02 & \\
			& 84\%&	12.02 &	9.21 &	10.82 &	44.59 &	15.12 & \\ \hline
			& 16\%&	3.47 &	8.80 &	10.74 &	40.22 &	0.00 & \\
QSOs	& 50\%&	7.87 &	9.32 &	10.92 &	42.38 &	0.09 & \\
			& 84\%&	12.11 &	9.64 &	11.05 &	44.28 &	7.46 &
\enddata
\end{deluxetable*}

\begin{figure*}[!htb]
\centering 
	\subfigure[]{ 
		\includegraphics[width=0.45\textwidth]{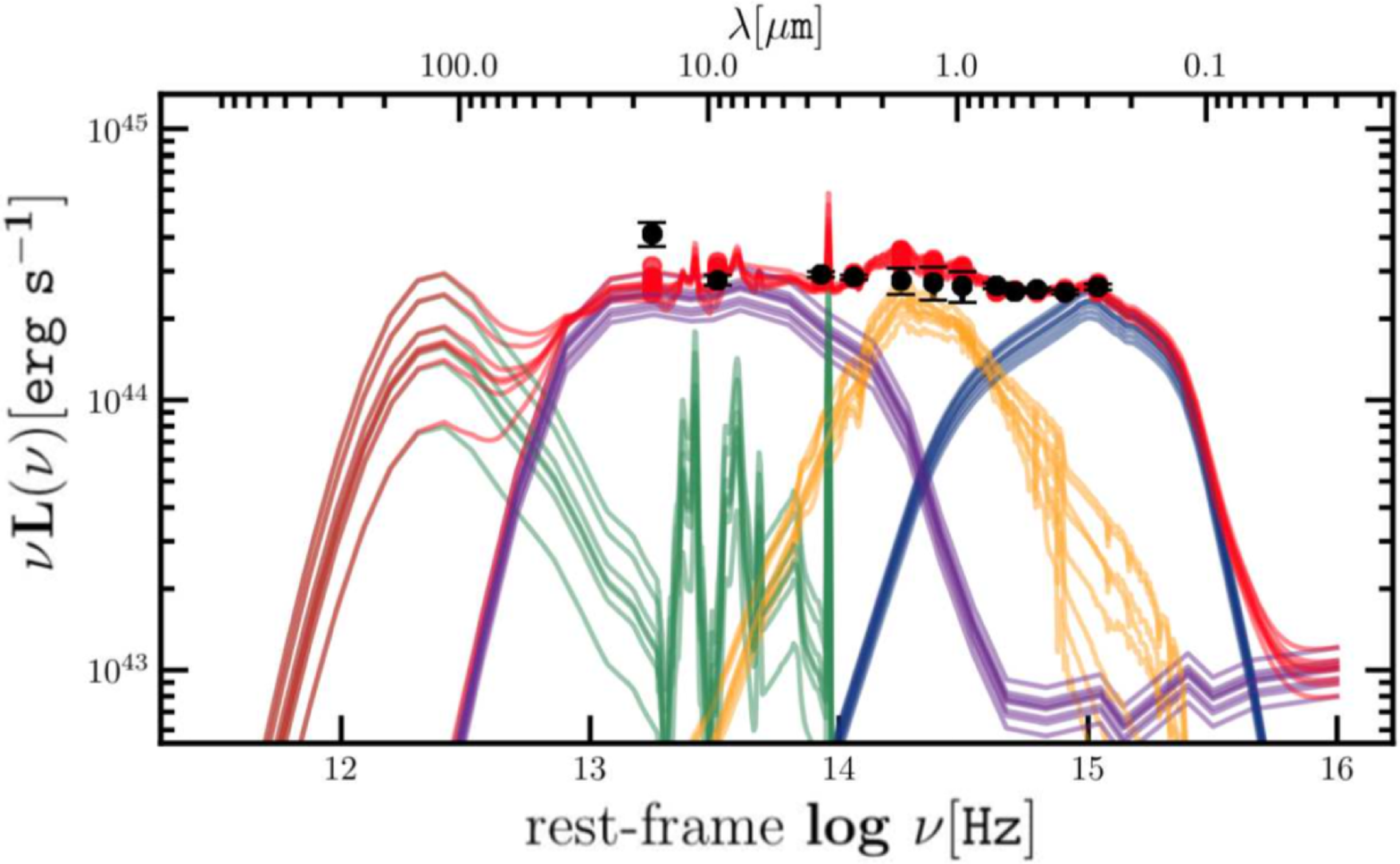}}
	\subfigure[]{ 
		\includegraphics[width=0.45\textwidth]{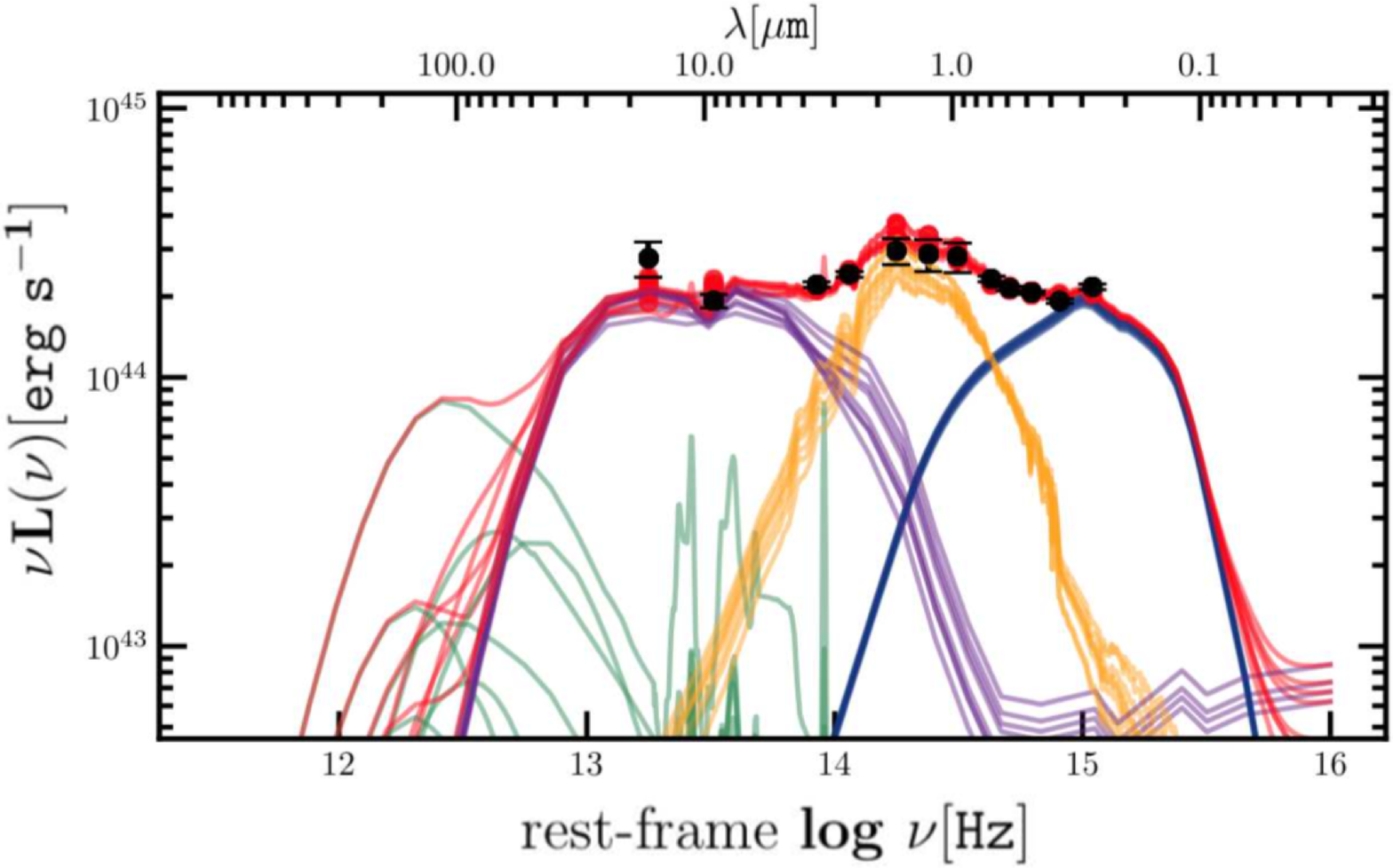}}
	\caption{\label{fig:sed_fitting} 
	SED fitting of the composite SED of PSQs (left) and QSOs (right): The black 
	circular markers with error bars indicate the observed photometric data. The 
	other lines represent the individual model components, namely, starburst 
	component (green), host galaxy component (orange), the hot-dust emission 
	(purple) and the BBB  template (blue), while the red line shows the linear 
	combination of these, i.e., the 'total SED'. Eight different realizations 
	which are picked randomly from the parameter posterior PDFs are plotted to 
	show the uncertainties of the parameters on the SEDs.}
\end{figure*}

\autoref{fig:sed_fitting} shows the composite SEDs and the AGNfitter fits. We scaled the 
 composite SEDs to the median 4.6\micron\ luminosity of the corresponding sample. The 
 starburst (green), host galaxy (orange), the hot-dust emission (purple) and the 
 ``big-blue bump'' (blue) components combine for the composite model (red), which is 
 shown in the panels in \autoref{fig:sed_fitting}. We randomly pick eight different 
 realizations from the posterior PDFs and over-plot the corresponding component SEDs 
 in order to visualize the dynamic range of the parameter values included in the PDF. 
 In the \autoref{tab:table2} and \autoref{fig:sed_fitting}, we can see our composite 
  PSQ SED shows younger stellar age, shorter exponential star formation history 
 (SFH) timescale $\tau$, and higher specific star formation rates (sSFR; SFR/\Mstar) 
 than those of the QSOs sample. The fitting result of age $\sim$ 500 Myr supports 
 that our PSQs sample have the intermediate-age hosts.

\section{Summary}\label{sec:summary}

We constructed a catalog of 208 post­-starburst quasars (PSQs) from the SDSS DR7
quasar catalog. The PSQs have  \hd\ absorption equivalent width EW(\hdhost)
{greater than or equal to} 6 $\AA$, indicating that they contain large fraction
of intermediate-­aged stellar populations formed in a recent burst of star
formation. Our catalog will be used to study the relationship between
post­-starburst quasars and normal quasars. In particular, in our future work,
we will test the hypothesis that post­-starburst quasars represent an important
phase in the evolution of the massive galaxies and that supermassive black holes
and their hosts co-­evolve. To that end, we will study additional properties of
the PSQs and their host galaxies such as their ages, stellar masses, SFRs, black
hole masses and Eddington ratios.

\acknowledgments

The authors are very grateful to the anonymous referee for critical comments and
instructive suggestions, which significantly strengthened the analyses in this
work. We thank Z. Shang for valuable discussions. This work is supported by the
National Key Research and Development Program of China (No. 2017YFA0402703 and
2017YFA0402704), and by the National Natural Science Foundation of China (Nos.
11733002 and 11673057).

Funding for the SDSS and SDSS-II has been provided by the Alfred P. Sloan
Foundation, the Participating Institutions, the National Science Foundation, the
U.S. Department of Energy, the National Aeronautics and Space Administration,
the Japanese Monbukagakusho, the Max Planck Society, and the Higher Education
Funding Council for England. The SDSS Web Site is http://www.sdss.org/.

The SDSS is managed by the Astrophysical Research Consortium for the
Participating Institutions. The Participating Institutions are the American
Museum of Natural History, Astrophysical Institute Potsdam, University of Basel,
University of Cambridge, Case Western Reserve University, University of Chicago,
Drexel University, Fermilab, the Institute for Advanced Study, the Japan
Participation Group, Johns Hopkins University, the Joint Institute for Nuclear
Astrophysics, the Kavli Institute for Particle Astrophysics and Cosmology, the
Korean Scientist Group, the Chinese Academy of Sciences (LAMOST), Los Alamos
National Laboratory, the Max-Planck-Institute for Astronomy (MPIA), the Max-
Planck-Institute for Astrophysics (MPA), New Mexico State University, Ohio State
University, University of Pittsburgh, University of Portsmouth, Princeton
University, the United States Naval Observatory, and the University of
Washington.

\appendix
\section{Appendix material}
\subsection{Mock spectra test}\label{sec:test}

\begin{figure}[!htb]
\centering 
	\center{\includegraphics[width=0.8\textwidth]{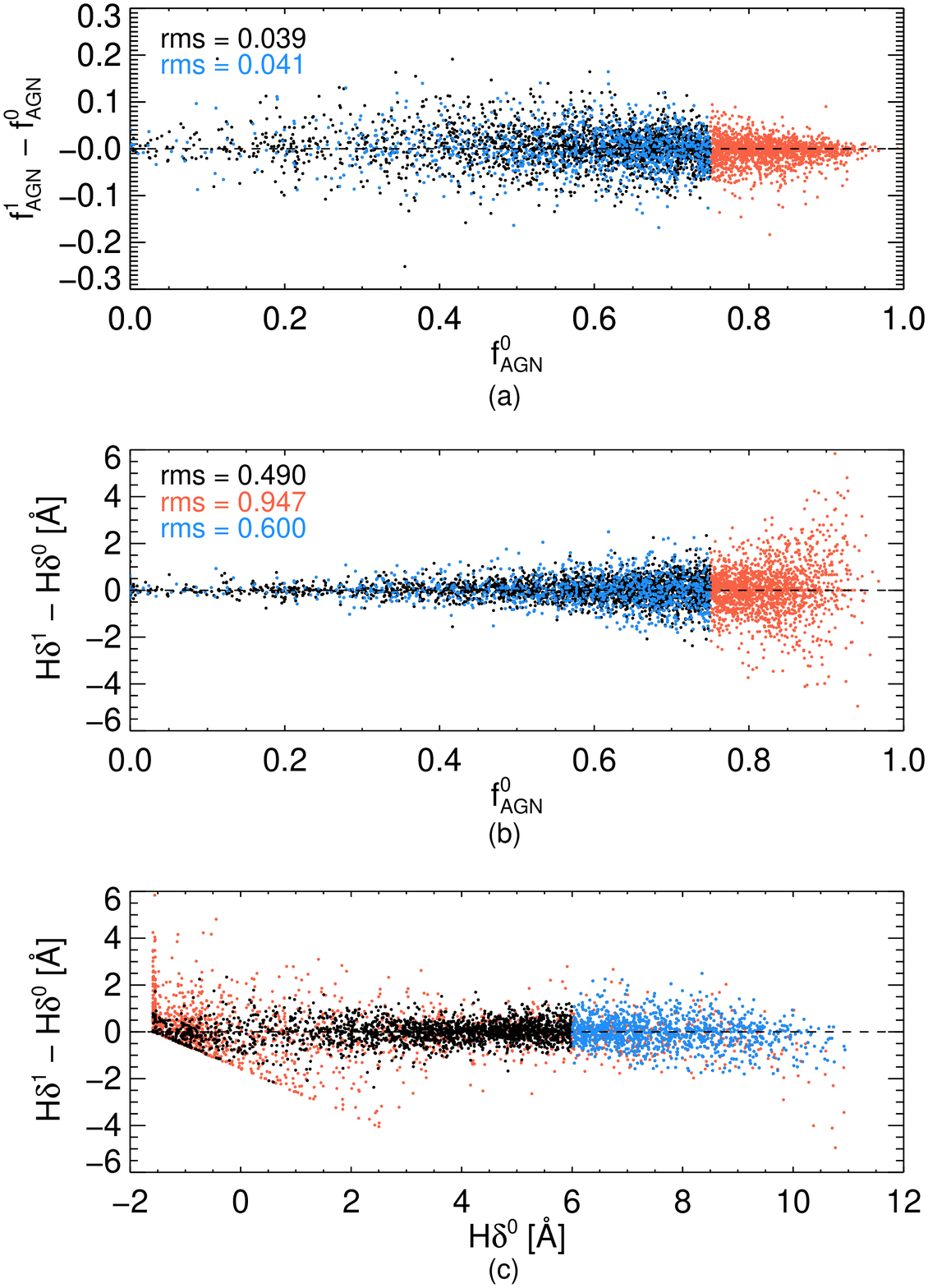}}
	\caption{\label{fig:test} 
Comparison between the input parameter with the difference of the output from 
input parameter (a) input AGN fraction and the different between output and 
input AGN fraction, (b) input AGN fraction and the different between output 
and input \hdhost, (c) input \hdhost and the different between output and 
input \hdhost. The black dots represent the spectra with reliable fits of the 
stellar component (output AGN fraction \fagn $<$ 0.75), while the blue dots 
represent those with post-starburst signature \fagn $<$ 0.75  and \hdhost $>$ 
6 \AA). The red dots show the remaining spectra with poor fits (\fagn $>$ 0.75). 
The dash lines represent the identity lines. The numbers in those panels indicate 
the RMS scatter around the identity line, its colors are corresponding to the 
colors of dots.}
\end{figure}

In order to evaluate the reliability of our method, we generated 5000 mock
spectra by combining spectral models of quasar power-law continuum, Fe emission 
template, broad Balmer emissions and 6 SSPs host stellar population as described 
in \autoref{sec:fitting}, and 10\% random noise. We built a probability distribution 
function (PDF) of each input parameter by a linear interpolation of the measured 
distribution of that parameter. The fitting results of the free parameters are 
plotted in \autoref{fig:test}. We found that the output best-fitting \fagn\ is 
reliable, and the scatter of EW(\hdhost) is not greater than 0.6\AA\ until the AGN 
comes to dominate the spectral flux (\fagn $> 75\%$). We therefore excluded the objects 
with \fagn $> 75\%$ which their host results are considered unreliable. In Figure 
A1 (c), there are some patterns in the the region between -2\AA\ and 2\AA. The 
reason is that when the Balmer absorption features are not obvious,  e.g. the 
age of host is relatively old, our fitting result may reach the boundary value 
of -2\AA. But this problem of an old age host does not affect the results of 
this investigation. 

\begin{figure}[!htb]
	\center{\includegraphics[width=0.75\textwidth]{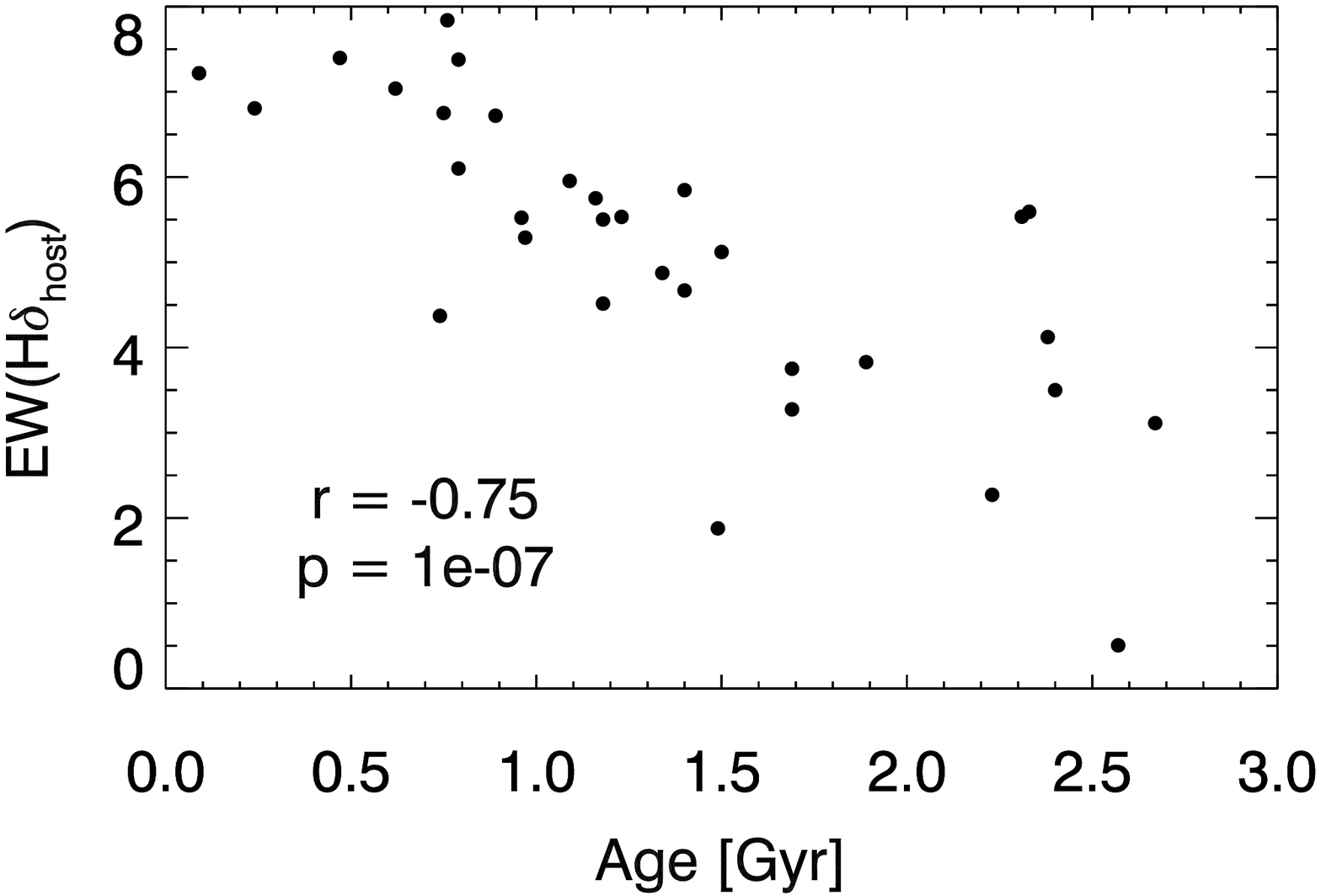}}
	\caption{\label{fig:c13_test} 
	The EW(\hdhost) measurements plotted against the starburst ages of PSQs
	from C13. There is a linear correlation with a Pearson r value of $-$0.75,
	and the p-value is 1e$-$7.}
\end{figure}

We also applied our spectral-fitting method for the sample of C13 to further test our method. 
Although our sample and theirs only partially overlap because the AGN luminosity and post-starburst 
criteria are different, we still obtained a tight correlation between our
EW(\hdhost) measurements and the starburst ages of PSQs from C13 (see \autoref{fig:c13_test}), showing that
our methods are at least consistent with this previous careful work. 

\bibliographystyle{aasjournal}
\bibliography{wei16}

\clearpage

\end{document}